\documentclass[lettersize,journal]{IEEEtran}
\usepackage{amsmath,amsfonts}
\usepackage{algorithmic}
\usepackage{algorithm}
\usepackage{array}
\usepackage[caption=false,font=normalsize,labelfont=sf,textfont=sf]{subfig}
\usepackage{textcomp}
\usepackage{stfloats}
\usepackage{url}
\usepackage{verbatim}
\usepackage{graphicx}
\usepackage{cite}
\begin{document}

\title{Photonic Generation and Free-Space Distribution of \\ Millimeter Waves for Portable Optical Clocks }

\author{Dylan Meyer, Alexander Lind, William Groman, Hero Trent, Carter Mashburn, Matthew Heyrich, \\Jeffrey Sherman, Franklyn Quinlan, Gabriel Santamaria-Botello, Scott A. Diddams

\thanks{This work has been submitted to the IEEE for possible publication. Copyright may be transferred without notice, after which this version may no longer be accessible. Dylan Meyer, Alexander Lind, William Groman, Franklyn Quinlan, and Scott Diddams are affiliated with the University of Colorado, Boulder, Electrical, Computer, and Energy Engineering Department, 1111 Engineering Dr, Boulder, CO 80309. William Groman, Hero Trent, Carter Mashburn, Matthew Heyrich and Scott Diddams are affiliated with the University of Colorado, Boulder, Physics Department, 390 UCB University of Colorado Boulder, CO 80309. Jeffrey Sherman and Franklyn Quinlan are affiliated with the National Institute of Standards and Technology, 325 Broadway, Boulder, CO 80305. Gabriel Santamaria-Botello is affiliated with the Colorado School of Mines Department of Electrical Engineering, Brown Hall 310C, 1610 Illinois St., Golden, CO 80401.}}

\maketitle

\begin{abstract}
Robust and portable optical clocks promise to bring sub-picosecond timing instability to smaller form factors, offering possible performance improvements and new scenarios for positioning and navigation, radar technologies, and experiments probing fundamental physics. However, there are currently limited methods suitable for broadly disseminating the sub-picosecond timing signals or performing frequency comparison of these clocks--particularly over open-air paths. Established microwave time transfer techniques only offer nanosecond level time synchronization, whereas optical techniques have challenging pointing requirements and lack the capability of all-weather operation. In this paper, we explore optically derived millimeter-wave carriers as a time-frequency link for full utilization of the next generation of portable optical clocks. We introduce an architecture that synthesizes 90 GHz millimeter waves with a one second residual instability of $2\times10^{-15}$, averaging into the $10^{-17}$ range. In addition, we demonstrate a first-of-its-kind 110 m phase-stabilized free-space frequency comparison link over a millimeter-wave band with a one second instability in the $10^{-14}$ region. Technical and systematic uncertainties are investigated and characterized, providing a foundation for future time and frequency transfer experiments among distributed portable optical clocks.
\end{abstract}


\section{Introduction}
\IEEEPARstart{S}{ince} their initial realization in the early twenty-first century, optical clocks have progressed from a proof of concept to serving as the benchmark for low noise and high stability timekeeping devices\cite{Ludlow}. The exquisite timing and frequency stability of these clocks promises to revolutionize navigation infrastructure, telecommunications, and cutting-edge probes of fundamental science \cite{Thorium,Katori,DM}. These optical clocks have continued to demonstrate increased reliability and robustness beyond the research lab, making the integration of these frequency standards into existing modern infrastructure a reality \cite{SeaClocks,OpticalEnsemble,AussieClock}. \\ 
 \indent While a clock operating in isolation can be a valuable tool, a powerful array of applications with  transportable optical clocks need a timing network to link the devices. This allows the generation of time-difference data, as required for precise positioning, communications network synchronization, or applications in multi-static radar and very-long-baseline imaging (VLBI). Currently, free-space methods in which distant optical clocks can be networked are insufficient. Optical time transfer preserves the femtosecond-level precision of optical frequency standards, but is limited by requirements for precise pointing and clear line-of-sight atmospheric conditions \cite{OTFT,Gozz,VAOTFT,T2L2}.
 \begin{figure}[!t]
\centerline{\includegraphics[width=\columnwidth]{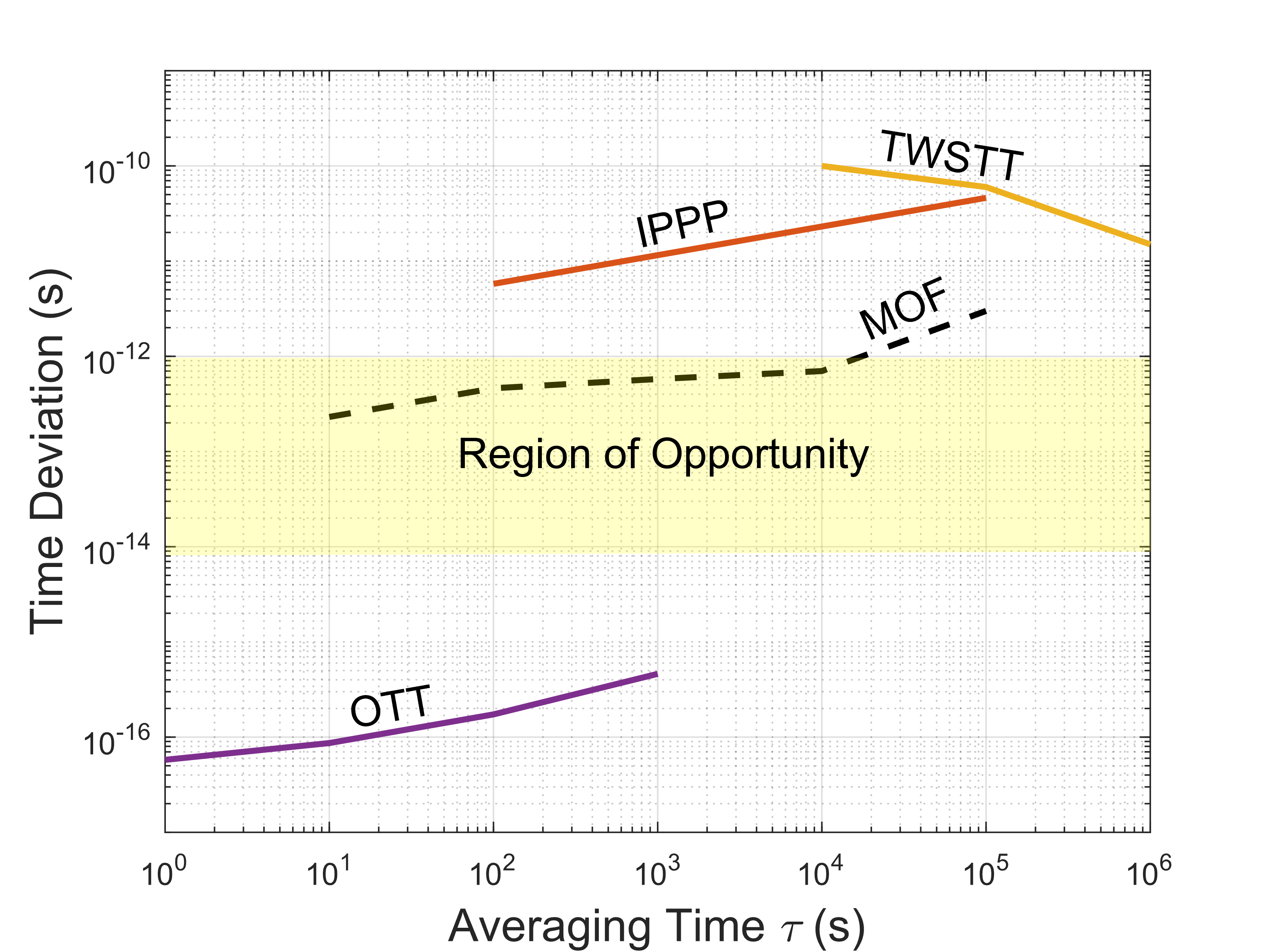}}
\caption{Time deviation of various time transfer techniques. Shown in the yellow region is a gap between flexible-pointing microwave techniques and optical free-space time transfer (OTT), the target region for millimeter wave based time and frequency transfer. In red, Integer Precise Point Positioning (IPPP), a GNSS based time transfer technique, is shown \cite{IPPP}. In yellow, two-way satellite time transfer (TWSTT) is shown \cite{TWSTT}. For context, a fiber based microwave over fiber (MOF) time distribution network is represented in black, but this approach is exclusively point-to-point for stationary sites \cite{ELSTAB}. The purple trace shows the stability of OTT \cite{OTFT}. Time deviation is used as these techniques are time transfer protocols with encoded timing information, not just frequency comparison techniques.}
\label{fig1}
\end{figure}
On the other hand, microwave time transfer techniques are more robust and easier to implement than optical time transfer, but suffer in performance, requiring excessively long integration times to support the performance of optical clocks\cite{SatFreq,IPPP ,TWSTT,ELSTAB}. While bespoke solutions such as the microwave link for the Atomic Clock Ensemble in Space (ACES) mission's aim to access these instability regimes~\cite{ACES}, this gap in the time transfer trade space is largely unexplored, especially when considering compatibility with optical clock systems.
This trade space is illustrated further in Fig. \ref{fig1}, where the time deviation for various time transfer methods is compared. As illustrated in the figure, there is a large gap in the range of $10^{-12}$ s to $10^{-14}$ s (10 fs to 1 ps) region of timing deviation for flexible-pointing time-transfer solutions. This is the same region of stability required to utilize new portable optical clocks to their fullest potential. In addition to meeting these performance standards, a time and frequency transfer system to support integration of optical clocks into existing infrastructure and other scientific or engineering use cases needs to reliably operate in all environments over kilometer lengths with flexible pointing. Figure \ref{ConceptFig} illustrates platforms that could host portable optical clocks that would benefit from an envisioned millimeter wave clock network.

\begin{figure}[!t]
\centerline{\includegraphics[width=\columnwidth]{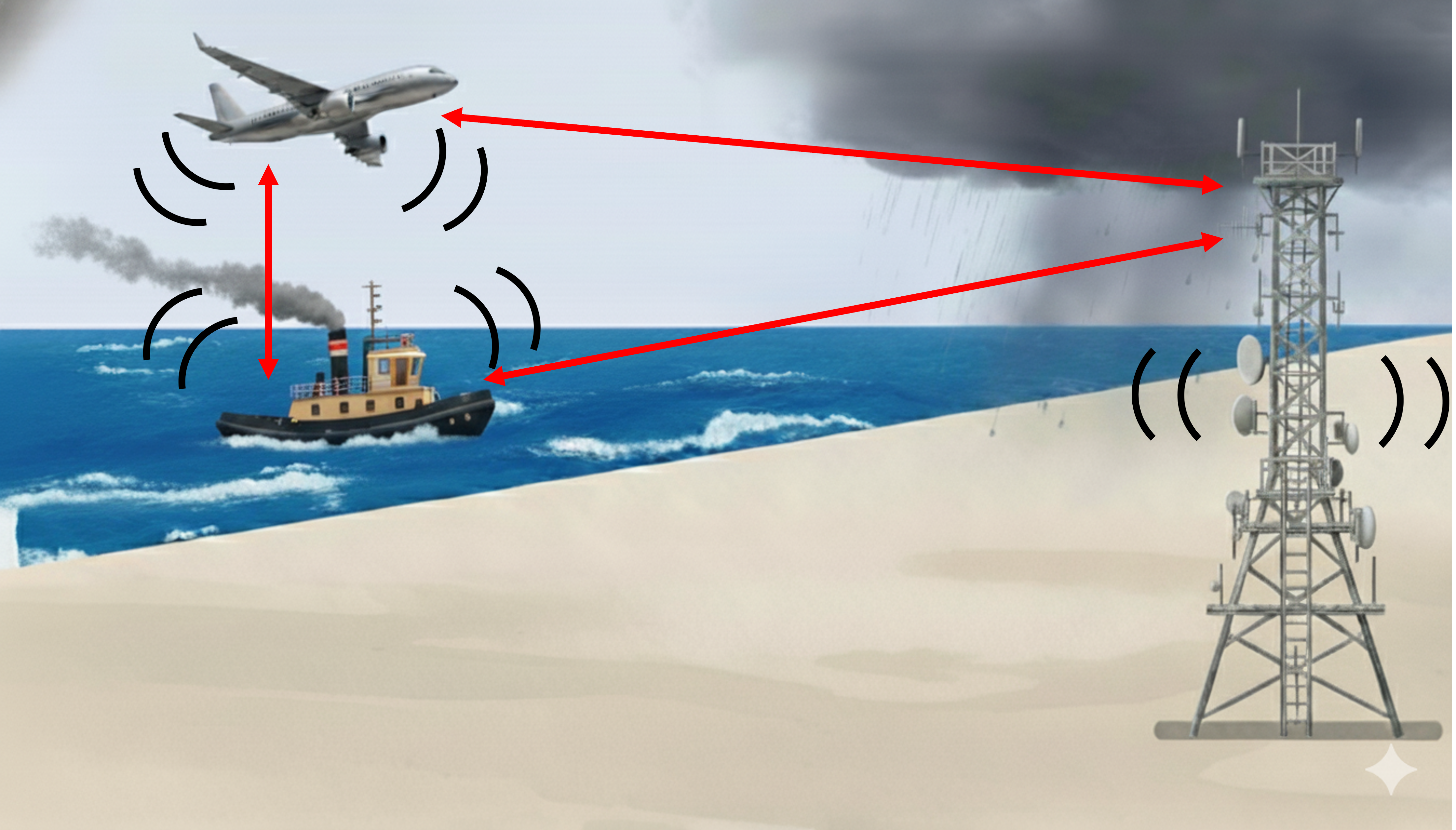}}
\caption{An illustration of an envisioned millimeter wave time transfer network. Mobile and stationary platforms, each with an atomic clock, would be capable of one or two-way time transfer over kilometer scale links through turbulent weather.\cite{Gem}}
\label{ConceptFig}
\end{figure}
\indent An alternative to traditional RF or optical time-transfer techniques is the use of millimeter-wave carriers for high-fidelity time-and-frequency transfer. Compared to RF systems at 1-10 GHz, the high frequency of the millimeter-wave carriers, typically in the range of 30-300 GHz, offers better timing discrimination.  At the same time, millimeter wave approaches have the benefits of fewer limitations in pointing and environmental operation when compared to optical time transfer\cite{km}. Furthermore, the growth of the millimeter wave communications market is providing better and lower cost components that lend themselves well to the implementation of robust and simple techniques used in microwave time transfer.\\ 
\indent Millimeter-wave bands have already been used to disseminate timing information through an upconverted radio frequency oscillator \cite{APL}. 
However, upconversion of RF oscillators adds multiplicative phase noise, degrading the performance of the microwave oscillator being transferred. In contrast to RF upconversion, directly generating millimeter waves via heterodyne detection or optical frequency division on a fast photodetector offers a phase coherent link from an optical clock laser, directly translating its stability to the millimeter wave domain\cite{IMRA,KAIST, Fortier,Radar}. 

While stable optical millimeter wave generation has been demonstrated through various techniques in the laboratory, the effect of free-space propagation on the stability of millimeter waves has not been studied\cite{Sweeper,AWGs,OEGen}. This is a critical step in establishing the legitimacy and performance of this proposed technique, as well as proving increased technology readiness levels that keep pace with portable optical clocks.
\\\indent In this work, we demonstrate a low residual noise millimeter wave generation technique capable of supporting the stability of portable optical clocks. Using these stable millimeter waves, we demonstrate a first-of-its-kind 90 GHz phase-stabilized free-space link over a 110 m folded path,  allowing for free-space millimeter wave frequency comparisons with fractional instability at the level of $10^{-14}$ at one second. To our knowledge this is the lowest uncertainty for the instability of such a millimeter wave link. As such, our millimeter wave generation technique and free-space dissemination represents a significant first step in the implementation of a stable free-space time and frequency transfer system with the capability of supporting portable optical clocks.

\section{Experimental Methods}

\begin{figure*}[!t]
\centerline{\includegraphics [width=6.5in]{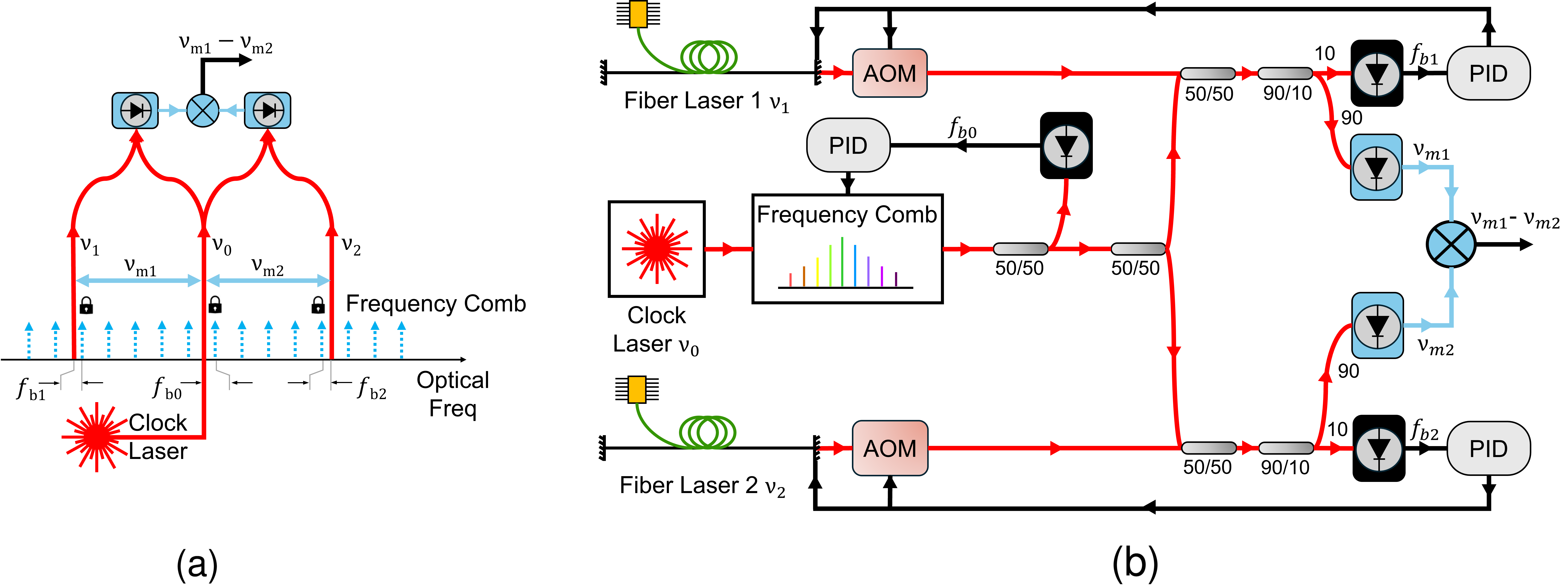}}
\caption{a. Frequency domain illustration of stable millimeter wave generation. The clock laser is represented as frequency $\nu_0$, while the frequency comb is represented by the blue dashed lines. Locking beatnotes are represented by $f_{b0}$, $f_{b1}$, and $f_{b2}$. The ancillary fiber lasers are represented as frequencies $\nu_1$ and $\nu_2$. Millimeter wave frequencies are determined by the spacing of the fiber laser and the clock laser, denoted as $\nu_{m1}$ and $\nu_{m2}$. b. Experimental schematic for stable millimeter wave generation and characterization. A clock laser at $\nu_{0}$ is used to stabilize an optical frequency comb. The light carrying the comb and clock laser is then split, and heterodyned with two fiber lasers. Via a heterodyne beat, these lasers are phase locked to the nearest comb tooth. The fiber laser and clock lasers are then heterodyned on high bandwidth photodetectors, which can then be mixed to characterize stability. Optical signals in fiber are denoted in red, millimeter wave signals and components in light blue, and radio frequency and baseband signals and components are shown in black.}
\label{Gen}
\end{figure*}
\subsection{Stable Millimeter Wave Generation}
Our approach uses an experimentally simple architecture that is compatible with any optical clock. We use three optical sources alongside auxiliary locking electronics: a cavity stabilized laser which serves as a substitute for an atomically disciplined clock laser, a frequency comb, and a telecom wavelength fiber laser. All optical clocks require a clock laser and frequency comb, meaning this technique only requires the addition of a continuous wave laser and locking electronics to generate stable millimeter waves from an optical clock output.  We note that our approach could also use recent advances in integrated photonics for implementation on a chip-scale photonic platform\cite{Igor,OFD,Groman}.\\
\indent Figure \ref{Gen}a illustrates the frequency domain representation of the generation scheme. A clock laser at $\nu_0$ is used to stabilize a self-referenced frequency comb through a radio frequency beatnote, $f_{b0}$, locking the comb mode spacing, $f_{rep}$, to match the stability of the clock laser \cite{FSLOC,Line}. A continuous wave laser is then stabilized to another comb mode that is offset by $\sim 90$ GHz via another beatnote $f_{b1}$. All parameters that would induce variation in the frequency spacing of the two lasers used in the millimeter wave heterodyne are stable, as the repetition rate of the comb, the carrier-offset-envelope frequency of the comb, and spacing of the continuous wave laser and the nearest comb tooth, are phase-locked. The clock laser and continuous wave fiber laser are then heterodyned to create a microwave or millimeter wave tone at an arbitrary difference frequency of the two lasers, $\nu_{m1}$. To characterize the relative stability of this technique and subsequent millimeter wave propagation experiments, this setup is duplicated to generate two different millimeter waves from the same common optical source and comb. The exact frequencies of the two millimeter waves are slightly offset and can then be heterodyned to generate an RF tone for analysis with laboratory electronics. \\
\indent Fig. \ref{Gen}b illustrates the component-level setup of the millimeter wave generation experiment. A 100 MHz self-referenced erbium-fiber frequency comb is stabilized through a heterodyne beat with the cavity-stabilized laser using a Proportional-Integral-Derivative (PID) control scheme, fixing the frequencies of all modes of the frequency comb relative to the stable laser.  The light from this stabilized laser and comb lock is split again in a 50/50 fiber coupler and heterodyned with another low-noise, tunable, 1550 nm fiber laser at $\nu_1$. This mixed light is then split in a 90/10 fiber coupler, with 90 percent used to generate a millimeter wave tone and 10 percent used for locking the fiber laser. While comb light is present on the millimeter wave photodetector, the tones generated by the continuous wave lasers heterodyning with the comb modes are weak when compared to the 90 GHz carrier. In addition, the repetition rate harmonics are weak compared to the 90 GHz carrier. Therefore, these contributions can be ignored. The PID feedback mechanism for the fiber laser locking to the nearest comb tooth is implemented through a split fast and slow lock. The fast lock actuator is an acousto-optic modulator (AOM) with a bandwidth of $\sim100$ kHz, and the slow lock is low-pass filtered at 10 Hz and then fed to a piezo-electric element in the  fiber laser to control the length of the fiber cavity. \\
\indent For this experiment, the fiber laser frequency $\nu_1$ is placed approximately 90 GHz away from the clock laser, such that  $\nu_{m1}=|\nu_0-\nu_1|$, as there is a favorable atmospheric transmission window in this band \cite{Al}. This process is repeated with a laser $\nu_2$ to produce a second millimeter wave at frequency $\nu_{m2}=|\nu_0-\nu_2|$. These two millimeter wave tones can be mixed, generating a tone at $\nu_{m1}-\nu_{m2}$. This beat is subsequently analyzed to quantify  the additive noise this technique imposes on an ideal optical clock output, as the noise present in the optical lock, $f_{b0}$, is common to both millimeter wave tones. 

\subsection{Free-Space Millimeter Wave Frequency Comparison}

{ Figure \ref{FreeSpace} shows the setup for the free-space 110 m folded W-band link. The free-space section of the experiment is conducted approximately 20 meters away from the optical table where the stable millimeter waves are generated. Transmit and receive setups are placed on carts separated by several meters in a hallway exterior to the lab. All optical tones for the receiver are sent to the receive station via a 20 m  fiber, where they are then demultiplexed using a fiber based dense wavelength division multiplexer filter. The transmit optical signals are sent via their own uncompensated fiber. The optical signals from the first millimeter wave generation scheme, $\nu_{m1}$, are split. One copy is passed to the transmit cart. The transmit arm of the experiment is capable of preactuating out phase fluctuations seen in the free-space link through feedback control. This preactuated signal is then sent over free-space via a 25dBi transmit antenna with an output power of 0.5 W. The signal is retroreflected and returned to a 25dBi receive antenna with low-noise amplifiers cascaded to provide 55 dB of gain.\\
\indent This received signal is then split with a 3dB waveguide coupler. One output is mixed to generate an error signal containing information of added phase fluctuations imposed by the link. The other output is sent to a second mixer for which the local oscillator (LO) input can be toggled between $\nu_{m1}$ and $\nu_{m2}$, respectively. When using $\nu_{m1}$ as the LO of the analysis mixer, an in-loop measurement of the link stabilization performance is made. Using $\nu_{m2}$ as the local oscillator of the second analysis mixer yields an out-of-loop characterization of the system. \\
\begin{figure}[!t]
\centerline{\includegraphics[width=\columnwidth]{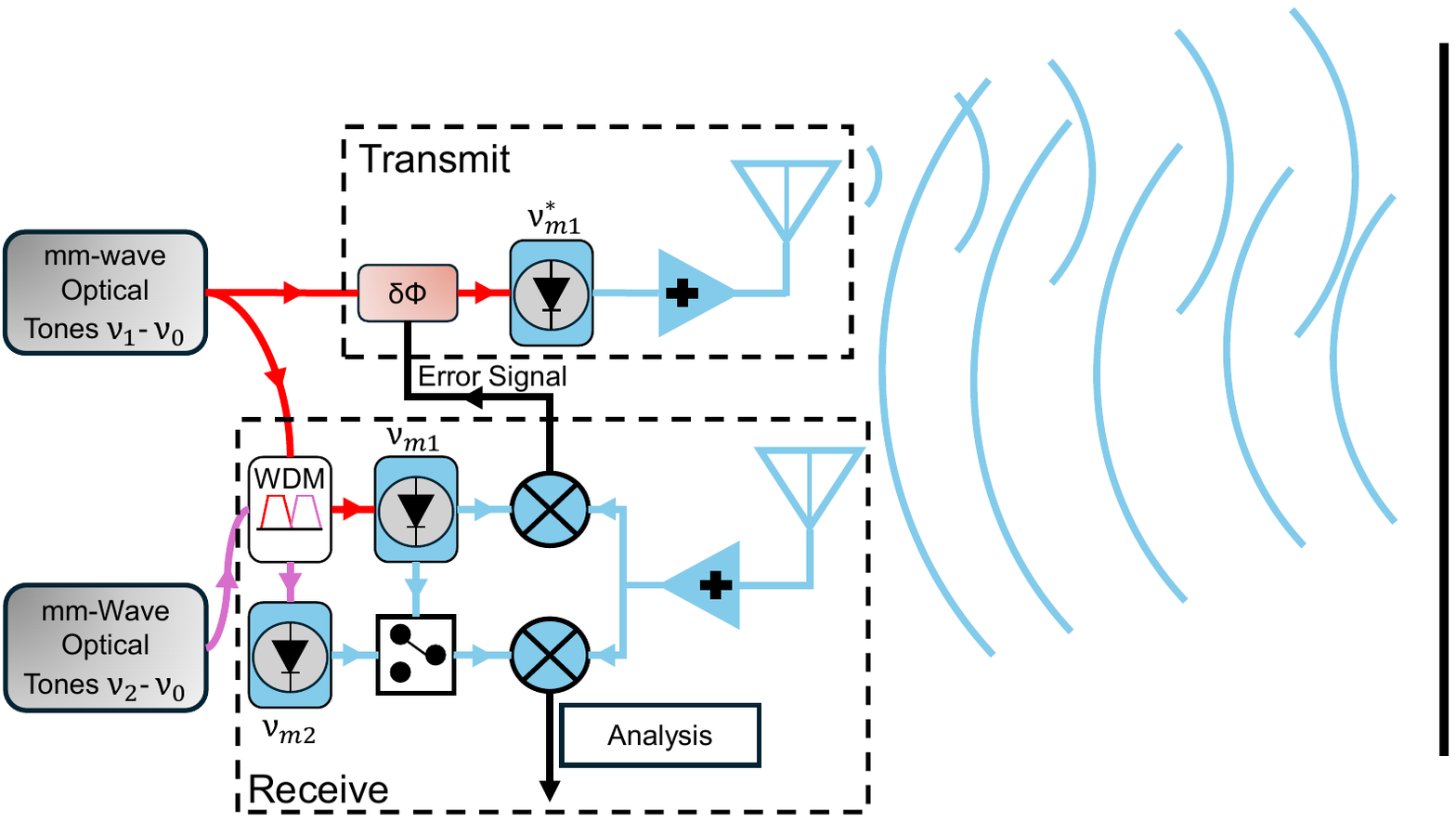}}
\caption{Diagram of free-space millimeter wave phase stabilized link. Red and pink lines are optical paths, blue lines are millimeter wave paths, and black lines are low-frequency electrical paths. Optical tones $\nu_0$ and $\nu_1$ are sent to a fiber splitter, with one copy being sent to the transmit and one to the receive carts. The transmit cart, capable of pre-actuating out path-induced phase fluctuations, generates $\nu_{m1}$, which is broadcast and retroreflected to the receive cart. The receive cart uses this signal to close the feedback loop to stabilized the link. Another mixer on the receive cart is driven by either $\nu_{m1}$ or $\nu_{m2}$, leading to in-loop or  out-of-loop analysis of the link performance.}
\label{FreeSpace}
\end{figure}
\section{Results and Discussion}
\subsection{Millimeter Wave Stability}
\begin{figure}[!t]
\centering
\subfloat[]{\includegraphics[width=3.5in]{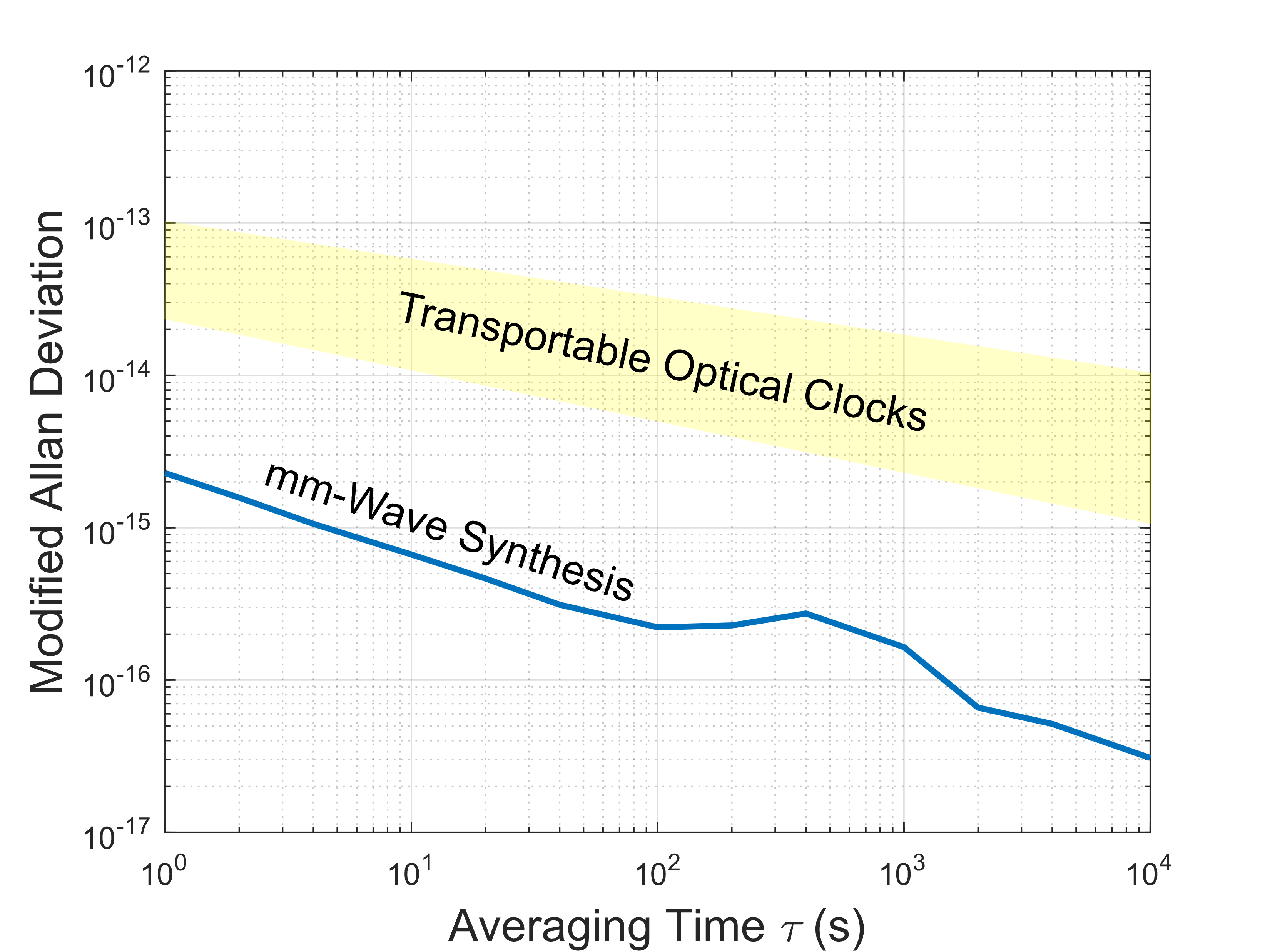}%
\label{mmwaveADEV}}
\hfil
\subfloat[]{\includegraphics[width=3.5in]{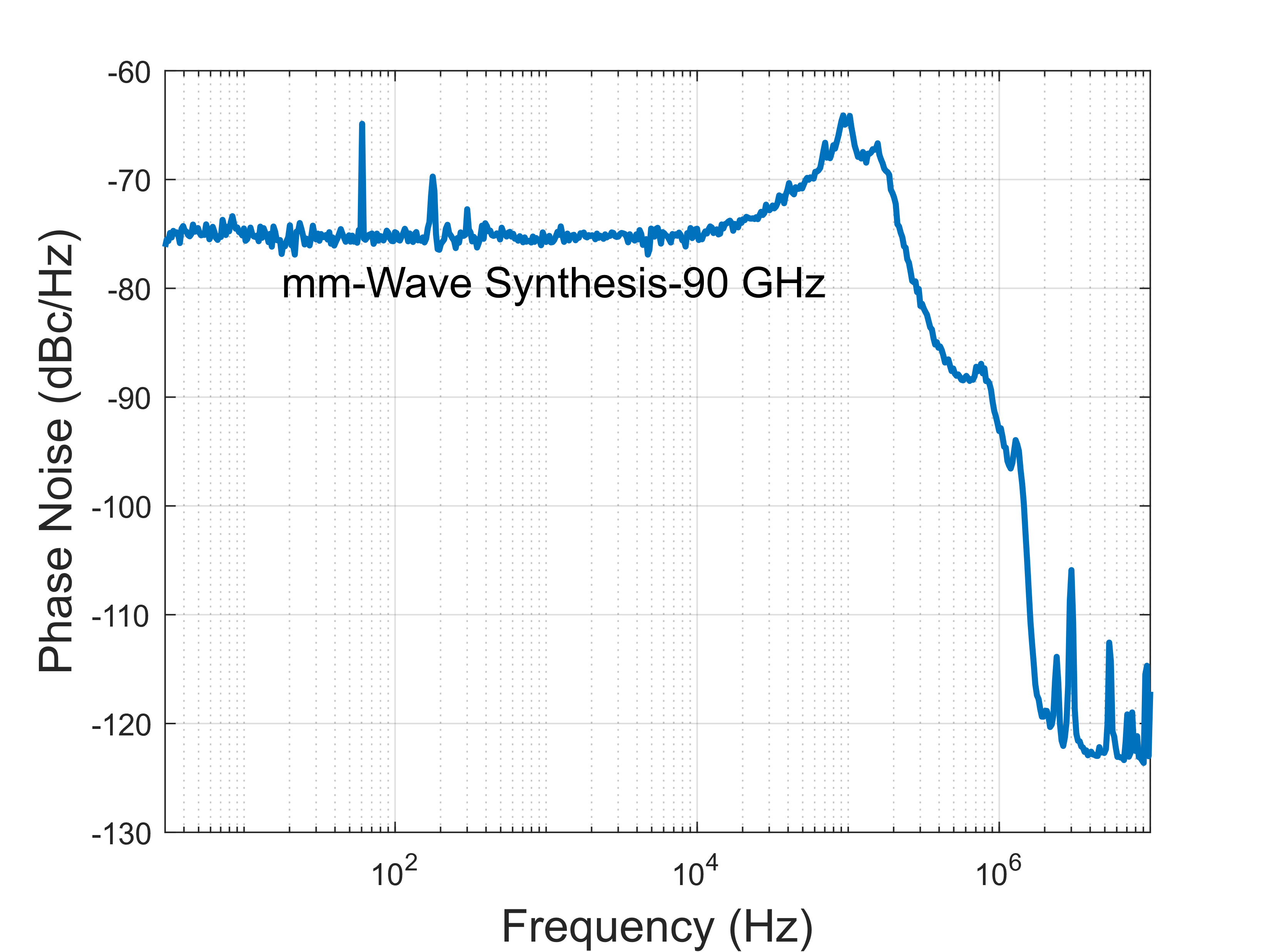}%
\label{mmwavePN}}
\caption{(a). MDEV of residual millimeter wave stability with the general transportable optical clock stability region seen in yellow. (b). Phase noise of millimeter wave beatnote. Transportable optical clock regions are defined by available and projected data sheets and papers from various companies and universities \cite{SeaClocks,AussieClock,VA,Vescent,Tiqker,Quantx,Turnkey}.}
\label{fig2}
\end{figure}
\indent The link is stabilized by an AOM modulating the fiber laser frequency $\nu_{1}$ used in the first millimeter wave generation scheme $\nu_{m1}$, frequency shifting the tone based on the error signal derived from the received retroreflected signal, yielding a new frequency $\nu_{m1}^*$. This error signal is generated by comparing the local reference copy of $\nu_{m1}$ to the received retroreflected signal that has propagated through turbulent air, revealing any noise added by the link. This error signal is then fed back to close the feedback loop. After the feedback loop is closed, in-loop and out-of-loop measurements are analyzed to gauge system stability. 

In a non-laboratory application, the error signal would need to be derived from a retroreflected signal measured at the transmit point, not the receive point. This can limit the achievable noise suppression for long links, but this demonstration still serves as a proof-of-concept for phase-stabilized millimeter-wave transmission\cite{PWilliams}.}\\ 
\indent To analyze the stability of the optically generated millimeter waves, the millimeter wave generation technique described in Section II is duplicated to generate a second millimeter wave tone. These two millimeter wave tones $\nu_{m1}^*$ and $\nu_{m2}$ are then mixed in a millimeter wave mixer, generating a radio frequency tone that can then be analyzed using laboratory electronics.\\
\indent Here we note that frequency stability metrics, as opposed to time stability metrics, are used for these experiments. This is because generating a stable frequency and disseminating this tone in free-space is the first step in full time-frequency link. In future experiments, we will add a time code on the millimeter wave carrier.\\
\indent For the stability measurement, an Agilent 53132A frequency counter is used to record a time series of frequency points, measuring the radio frequency tone over a measurement period with a gate time of one second. It should be noted that this is a $\Lambda$-type counter with millisecond level dead time. Therefore, modified Allan deviation (MDEV) is the best stability measure to use. More information on the effects of $\Lambda$-type counters, as well as the influence of dead time is given in\cite{Counter}. For phase noise analysis, a Fourier transform  spectrum analyzer with phase noise demodulation capabilities is used to measure offset frequencies of 3 Hz to 10 MHz.

\indent We begin by analyzing the beat $|\nu_{m1}-\nu_{m2}|$ as generated in Fig.~\ref{Gen}(b). The residual noise of the 90 GHz millimeter wave generation is quantified by the MDEV and phase noise as shown in Fig. \ref{fig2}(a) and Fig. \ref{fig2}(b), respectively. At one second integration time, the MDEV is $2\times 10^{-15}$, and averages down to the $10^{-17}$ level. At one second, the stability is limited by the in loop noise of the locks of $f_{b1}$ and $f_{b2}$, which both have an MDEV of $1.5\times 10^{-15}$ at one second. The slight increase in the 100s of seconds due to length fluctuations in the out-of-loop fiber optic cable that is caused by the cycling of the room temperature control. \\
\indent The phase noise of Fig. \ref{fig2}(b) is integrated from 3 Hz to 10 MHz to yield a timing jitter of 0.3 ps. The phase noise at offsets below 10 kHz is limited by the signal-to-noise ratio of the heterodyne beats $f_{b1}$ and $f_{b2}$ in Fig. \ref{Gen}. Around 100 kHz offset, we observe a servo bump due to the bandwidth limitations of the laser locking. And at offsets above 1 MHz, the signal-to-noise is  determined by the power of the millimeter waves $\nu_{m1}$ and $\nu_{m2}$ and their measured down-mixed signal.  \\
\indent It should again be noted that the clock laser and frequency comb noise are largely common to both millimeter wave tones, so these plots represent the residual stability of the millimeter wave generation technique. Essentially, this can be viewed as the noise floor for a millimeter wave generated using this technique if created by a perfect optical clock system, accounting for noise added by photodetection, electronic amplification, and servo control. As long as this stability remains above to the optical clock output, seen in Fig. \ref{fig2}(a), there will be negligible additive noise caused by the millimeter wave generation. Fig. \ref{fig2}(a) shows this technique is capable of supporting a range of portable optical clocks.
\subsection{Free-Space Link Stability}
The free-space link stability was first analyzed via an in-loop measurement. Monitoring the error signal used for link stabilization allows for the quantification of the noise floor of the stabilized link. To compare how well this stabilization technique works as opposed to a non-stabilized link, we compare MDEV and phase noise in two cases. The first case is where the feedback loop is closed, and the AOM is driven dynamically with the generated error signal as to mitigate noise. The second case is where the AOM is driven with a static 80 MHz tone from a low noise radio frequency synthesizer, and no feedback loop is closed, representing the stability of the path length itself.\\
\begin{figure}
\centerline\centering
\subfloat[]{\includegraphics[width=3.5in]{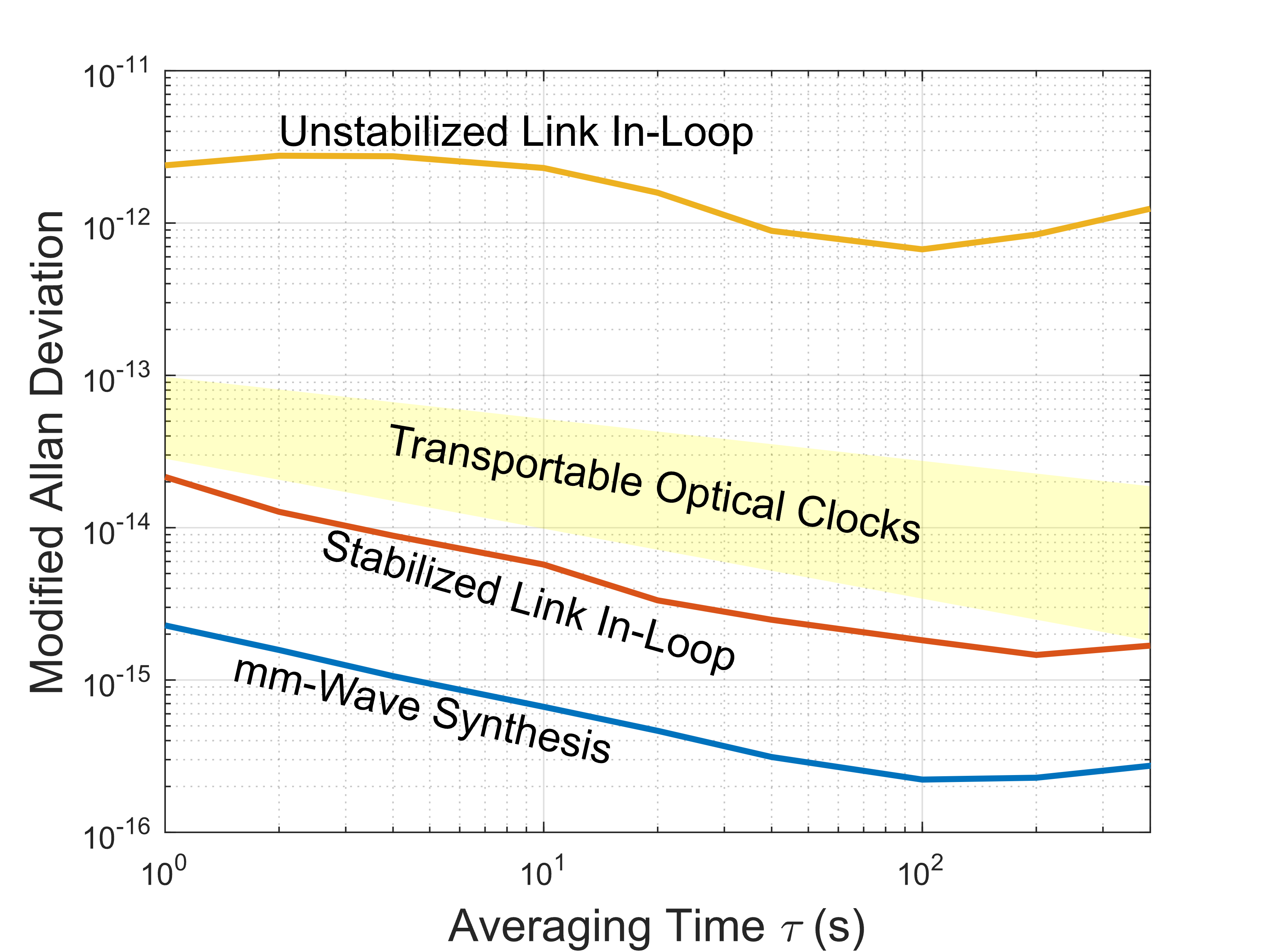}%
\label{InLoopADEV}}
\hfil
\subfloat[]{\includegraphics[width=3.5in]{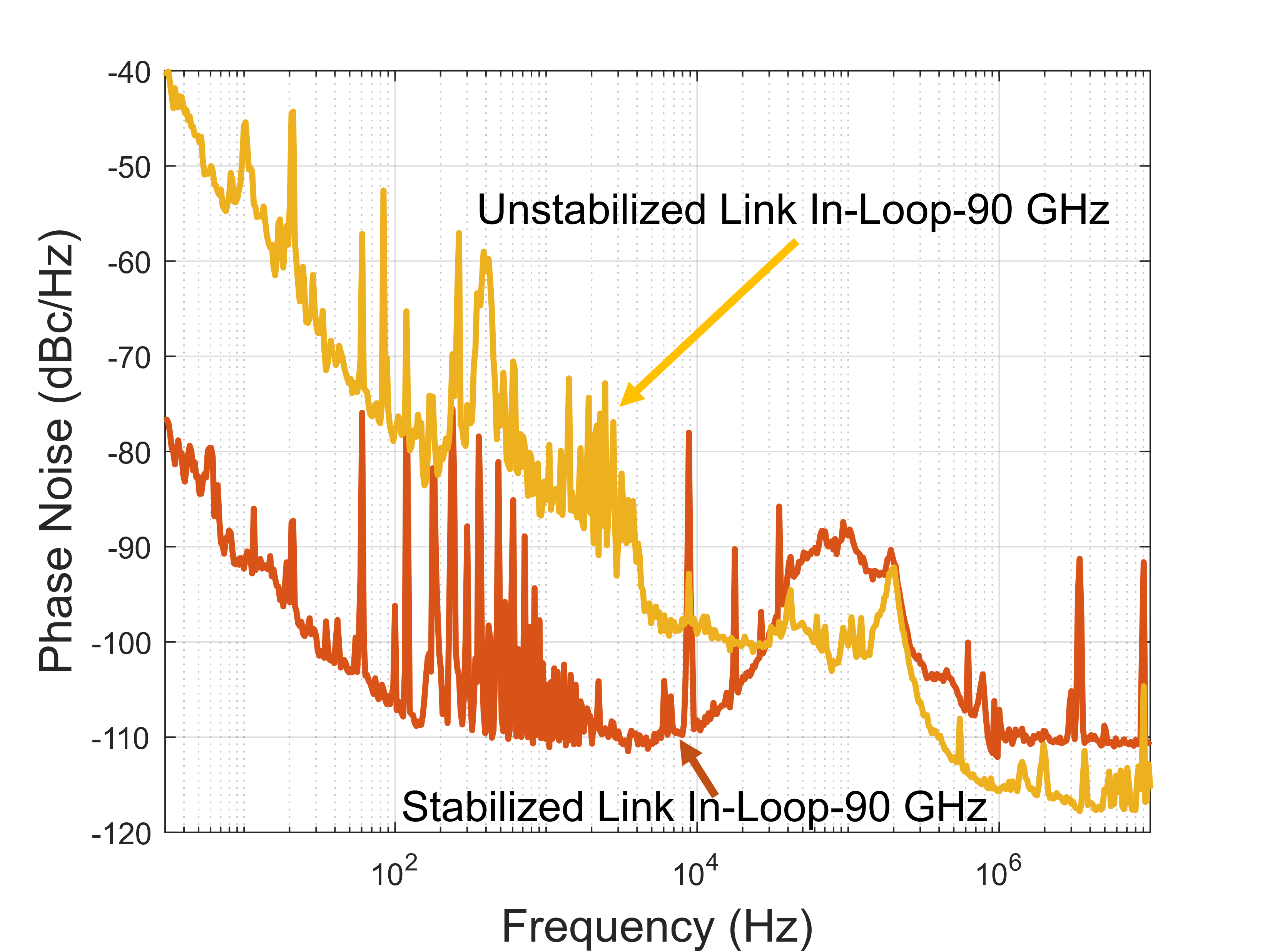}%
\label{InLoopPN}}
\caption{In-loop stability measurements of the phase stabilized free-space link. (a). in-loop  MDEV plots of unstabilized free-space link (yellow) and stabilized free-space link (red) alongside the performance region of portable optical clocks and the millimeter wave residual stability (blue). (b). Phase noise plots of in-loop unstabilized (yellow) and stabilized (red) free-space measurements. The unstabilized link represents the stability limit of the a free-space link without a feedback servo, while the stabilized link represents the stability limit of a free-space link with the servo engaged.}
\label{fig5}
\end{figure}
\indent Fig. \ref{fig5}(a) and Fig. \ref{fig5}(b) show the in-loop MDEV and phase noise of the in-loop measurements. These measurements are are measured with identical equipment and techniques as the millimeter wave stability measurements. As expected, closing the feedback loop greatly improves both phase noise and MDEV of the in-loop performance. The unstabilized link represents the noise of the unstabilized, free-space path due to a combination of index of refraction and density fluctuations, as well as the relative mechanical fluctuations of the transmit and receive terminals and the retroreflector. On the other hand, with the stabilized link the sum of all fluctuations are servoed out. Essentially, this would be the limit of achievable phase noise in the system if the link was limited by the servo. In reality, the millimeter wave phase noise is much higher in Fig. \ref{fig2}(b) than the stabilized link in-loop phase noise in Fig. \ref{fig5}(b), so the phase noise of the broadcasted millimeter wave is largely unaffected when the link is stabilized. \\
\indent In the case of MDEV, there are two orders of magnitude in improvement of the stabilized versus the unstabilized case, and approximately 40dB of improvement in low offset frequencies in the phase noise of the stabilized versus unstabilized case. Degradation of one second stability in comparison with the initial millimeter wave stability is to be expected. The increased instability is due to significantly higher noise caused by fluctuations in the link, the millimeter wave components being on carts with no vibration isolation, as well as loss caused by beam divergence and minor misalignment in the link, degrading the SNR due to 62 dB Friis antenna link loss. Despite these factors, the MDEV remains below the general range of portable optical clocks, showing that this technique is capable of supporting these systems for time and frequency transfer.\\
\indent For an out-of-loop measurement, the second millimeter wave $\nu_{m2}$ is compared against the pre-actuated transmitted stabilized link. Fig. \ref{fig6}(a) and Fig. \ref{fig6}(b) shows these stability results, with this stability trace shown in purple and labeled “comparison signal". We see that the short-term stability is comparable to the in-loop stability measurements, while the measured phase noise bears a strong resemblance to the previously measured phase noise of the directly measured millimeter waves. The deviation in the shape of the phase noise measurement of the free-space link versus the direct measurement of the in-waveguide millimeter waves is due to altered PID tuning parameters to allow for the system to reliably function on a non-vibrationally dampened cart, as well as link loss resulting in higher phase noise at offset frequencies above 1 MHz due to SNR degradation.\\
\indent The increase in MDEV at the 10's of seconds of the free-space comparison signal was investigated, and it was concluded this occurred as a result of thermal drift in uncompensated out-of-loop fiber. It should be noted in a field implementation of this technique, the fibers used to connect the optical heterodyne beats to the fast photodetectors could be stabilized using a Doppler canceled link similar to other works, which would allow the MDEV to continue to average down over longer averaging intervals \cite{KAIST}. \\
\indent In-loop measurements of the system show that this technique of free-space millimeter wave frequency comparisons is capable of supporting commercial optical clocks, while the demonstration of the free-space millimeter wave frequency comparison achieves an out-of-loop fiber limited measurement with excellent short term performance, which is then constrained by known out-of-loop systematics, predominantly out-of-loop thermal drift in optical fiber, over longer integration times.
\begin{figure}
\centerline\centering
\subfloat[]{\includegraphics[width=3.5in]{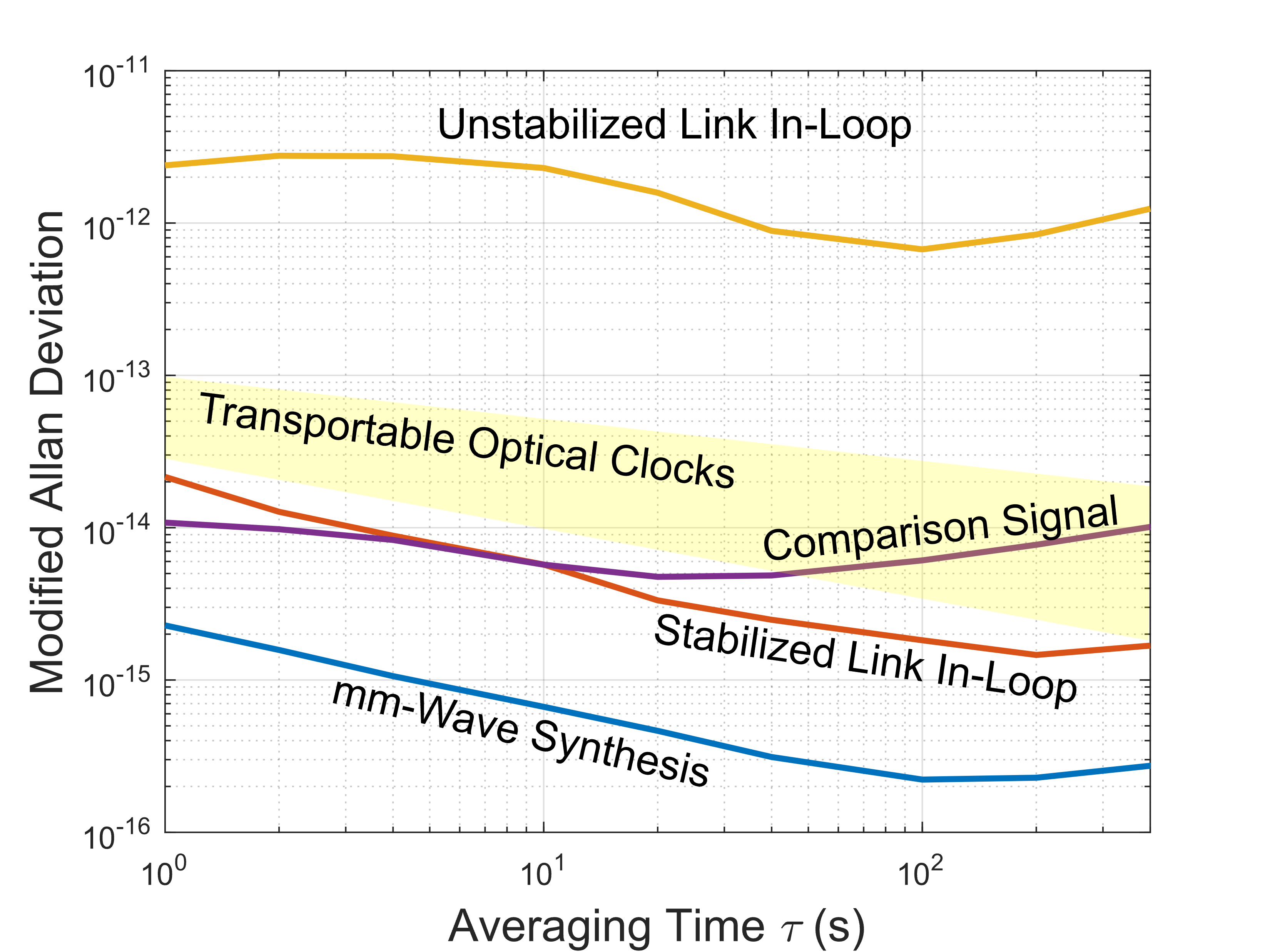}%
\label{OutOfLoopADEV}}
\hfil
\subfloat[]{\includegraphics[width=3.5in]{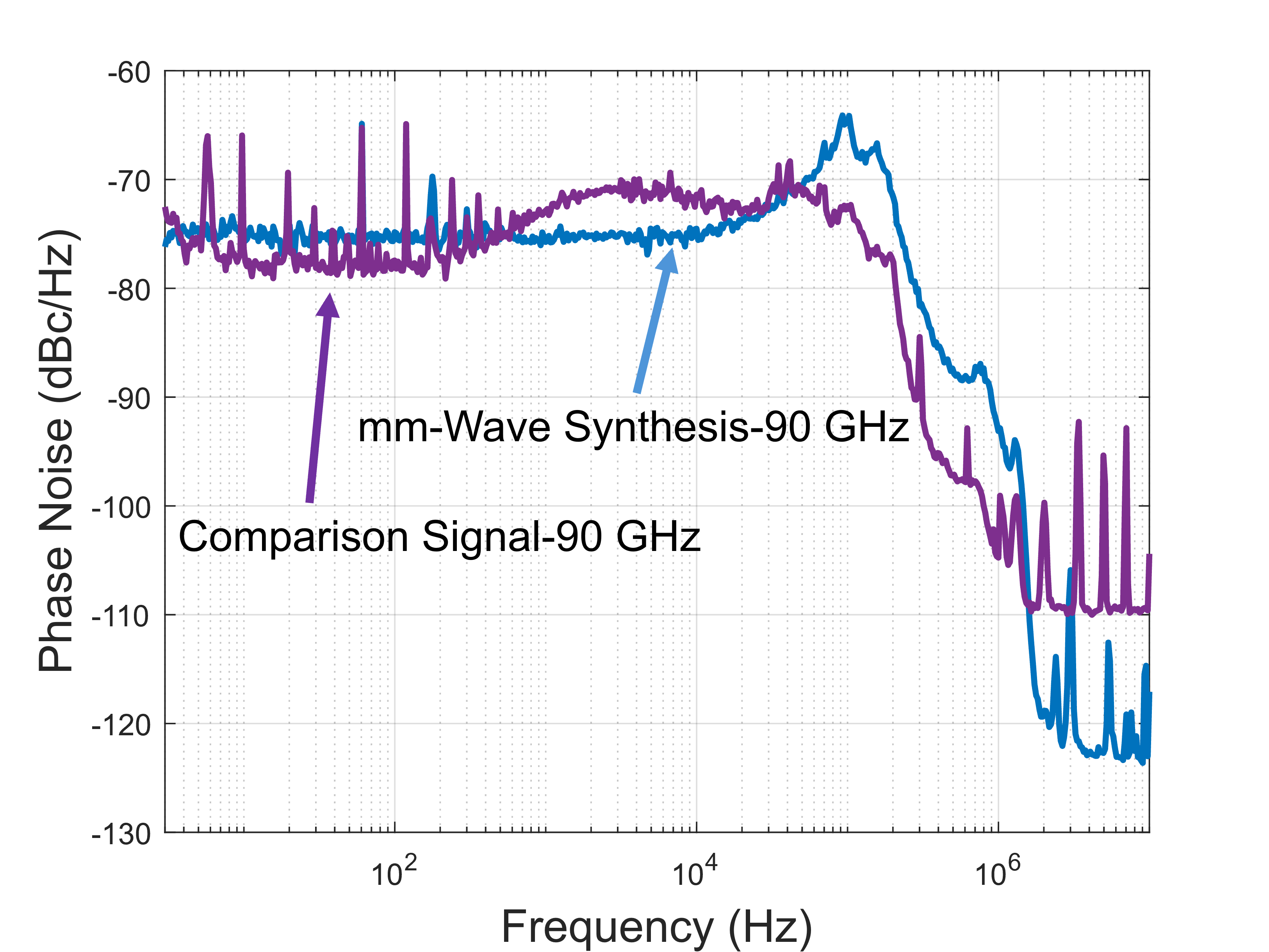}%
\label{OutOfLoopPN}}
\caption{Out-of-loop stability measurements of the phase stabilized free-space link. (a). MDEV of the out-of-loop free-space measurement, denoted as comparison signal, over a stabilized free-space link, with in-loop stabilized and unstabilized measurements. The  residual millimeter wave generation and transportable optical clock performances are also shown for context. The comparison signal drifts into the performance range of the transportable optical clocks due to known systematics, primarily out-of-loop temperature fluctuations in fiber. (b). Comparison of the phase noise of free-space millimeter wave measurement (purple), and the residual millimeter wave phase noise measurement (blue).}
\label{fig6}
\end{figure}
\section{Conclusion and Outlook}
We demonstrate a simple technique for generating optically derived millimeter waves with residual stability of $2\times10^{-15}$ at one second. In addition, we implement a method for distributing these millimeter waves over free-space for use in clock comparisons at the low $10^{-14}$ level at one second integration time. Limitations of the free-space comparison technique are identified as out-of-loop fiber optic paths experiencing thermal drift.\\
\indent Next steps for the improvement of this system would be the use of Doppler-canceled fiber links to mitigate mechanical and thermal fluctuations seen in out-of-loop fibers. Additionally, the implementation of a time code and phase measurements, allowing for the possibility of clock synchronization in addition to syntonization, is a logical next step. Implementation of these techniques over kilometer-scale links will offer an increasingly robust proof-of-concept of millimeter wave time and frequency transfer. This increase in length will allow for analysis of the coupling between link distance and stability performance.\\
\indent Implementing these techniques using integrated photonics could provide a platform for lower size, weight, and power (SWaP) stable time and frequency transfer. Similar two-tone based heterodyne techniques on integrated or low SWaP platforms have already been demonstrated with low phase noise\cite{Gryphon,Groman}. Investigating the long-term stability of these techniques, as well as the ability to implement high bandwidth actuation for use in a phase-stable link or the possibility of modulating a timing code on such platforms presents interesting possibilities and challenges for low SWaP time transfer supporting portable optical clocks.

\section{Acknowledgements}
This work is funded by the AFOSR FA9550-24-1-0142 and AFRL 282109-874X. D.M. and M.H. are partially supported by the NSF GRFP. The authors would like to thank Zoya Popovic and her lab members Stefan Stroessner and Alec Russell for their engaging conversation and assistance in W-Band power measurements.

\section{Appendix}
\subsection{Optimal Choice of Locking Frequencies to Reject Cavity Drift}
\begin{figure}[!t]
\centerline{\includegraphics[width=\columnwidth]{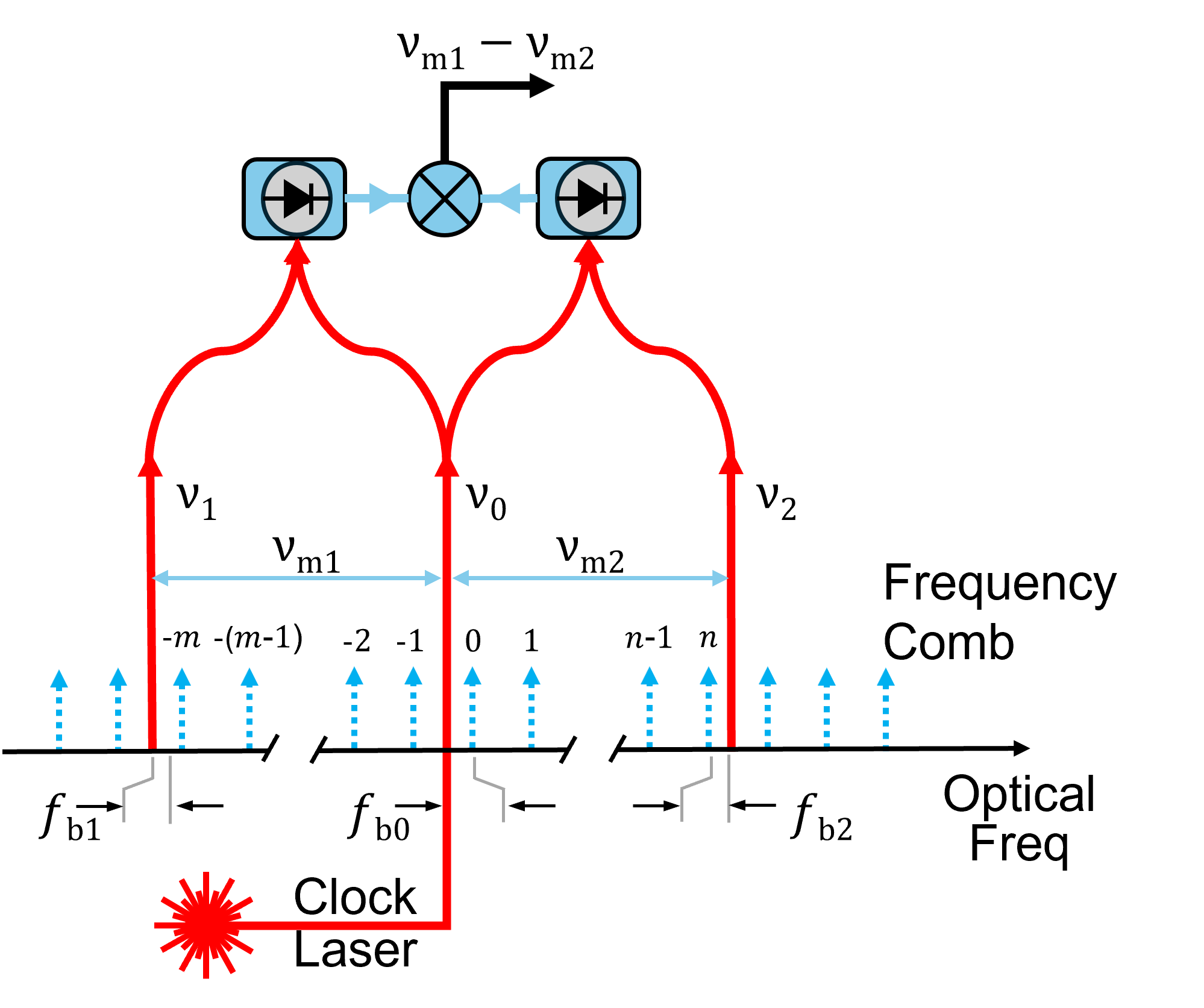}}
\caption{Detailed diagram of millimeter wave generation in frequency domain. }
\label{DriftRej}
\end{figure}
The placement of the two tunable 1550 nm fiber lasers on both sides of the cavity stabilized laser $\nu_0$ allows for rejection of frequency drift of the cavity, the comb repetition rate $f_{rep}$, and the microwave frequencies $\nu_{m1}$ and $\nu_{m2}$. However, this requires careful attention to the locking frequencies and exact comb modes employed.\\
\indent To fully reject any contribution of variations in $f_{rep}$ to the frequency instability of the millimeter wave generation technique, the tunable fiber lasers must be tied to equidistant comb modes away from the lock point of the comb to the cavity stabilized laser. As a result, if $\nu_0$ drifts , the repetition rate variation above and below the lock point are fractionally the same. This results in a change of spacing between the fiber lasers to the clock laser by $n$ comb modes multiplied by the repetition rate, where $n$ is the number of comb modes between $\nu_0$ and both $\nu_1$ and $\nu_2$. If the two fiber lasers are locked to $n$ and $m$ comb modes away from the clock laser respectively, there will be a component in the millimeter wave heterodyne frequency directly contributed to by the repetition rate multiplied by a factor of $m-n$. Since the cavity stabilized laser is not referenced to an atomic transition, the cavity frequency will drift. If $m-n$ is not $0$, then the millimeter wave frequency $\nu_{m1}-\nu_{m2}$ will also drift, even if the millimeter wave generation technique performs perfectly.\\
\indent To clearly illustrate this, an example millimeter wave generation experiment is shown in the frequency domain in Fig. \ref{DriftRej}. In the figure, $n$ and $m$ represent the number of comb modes away from the clock laser lock point. A general equation for $\nu_{m1}$ can be expressed in equation \ref{nu1}. An equation for $\nu_{m2}$ can similarly be expressed in equation \ref{nu2}.
\begin{equation}
\label{nu1}
\nu_{m1}=(f_{rep}-f_{b0})+(m-1)f_{rep}+f_{b1}
\end{equation}
\begin{equation}
\label{nu2}
\nu_{m2}=f_{b0}+nf_{rep}+f_{b2}
\end{equation}
Subtracting these two equations yields 
\begin{equation}
\label{diff}
\nu_{m1}-\nu_{m2}=-2f_{b0}+(m-n)f_{rep}+f_{b1}-f_{b2},
\end{equation}
which is a sum and difference of lock frequencies and the repetition rate, which is multiplied by the difference of $m$ and $n$. When $m=n$, the dependence on $f_{rep}$ is removed. In this case, as in the experiment, the remaining variables $f_{b0}$, $f_{b1}$, and $f_{b2}$ are all electronically locked to stable radio frequency references.  
It should be noted, if $f_{rep}$ is fixed by locking the clock laser to an atomic reference, this will not be an issue. So, while this was an important consideration for this experiment to properly analyze additive noise of the millimeter wave generation technique, the extra experimental steps relating to locking to the correct comb mode are not necessary when used when the underlying comb is locked to an optical frequency standard.

\end{document}